\documentclass[10pt,a4paper,onecolumn]{elsarticle}

\usepackage{graphicx}
\usepackage{color}
\usepackage{footnote}
\usepackage{epsfig}
\usepackage{natbib}
\usepackage{amsmath}
\usepackage{multirow}
\usepackage{hyperref}
\usepackage{bm}
\input{psfig.sty}
\makesavenoteenv{tabular}
\usepackage{fixfoot}
\newcommand{\angstrom}{\mbox{\normalfont\AA}}
\usepackage{algorithm}
\usepackage{algorithmic}

\begin{document}
\title{Predicting protein inter-residue contacts using composite likelihood maximization and deep learning}
\author[ICT,UCAS]{Haicang Zhang}
\author[ICT,UCAS]{Qi Zhang}
\author[ICT,UCAS]{Fusong Ju}
\author[ICT,UCAS]{Jianwei Zhu}
\author[ICT,UCAS]{Shiwei Sun}
\author[PKU1]{Yujuan Gao}
\author[HUST]{Ziwei Xie}
\author[PKU1]{Minghua Deng}
\author[ICT, UCAS]{Shiwei Sun$^{*,}$}
\author[ITP] {Wei-Mou Zheng$^{*,}$}
\author[ICT,UCAS]{Dongbo Bu$^{*,}$}

\address[ICT]{Key Lab of Intelligent Information Processing, Institute of Computing Technology, Chinese Academy of Sciences, Beijing, China}
\address[UCAS]{University of Chinese Academy of Sciences,  Beijing, China}
\address[PKU1]{Center for Quantitative Biology, School of Mathematical Sciences, Center for Statistical Sciences, Peking University, Beijing, China}
\address[ITP]{Institute of Theoretical Physics, Chinese Academy of Sciences, Beijing, China}
\address[HUST]{College of Life Science and Technology, Huazhong University of Science and Technology, Wuhan, China}
\cortext[cor1]{Correspondence should be addressed to W. Zheng (zheng@itp.ac.cn) and D. Bu (dbu@ict.ac.cn).}

\begin{abstract}
Accurate prediction of inter-residue contacts of a protein is important to calculating its tertiary structure. Analysis of co-evolutionary events among residues has been proved effective to inferring inter-residue contacts. The  Markov random field (MRF) technique, although being widely used  for contact prediction, suffers from the following dilemma: the actual likelihood function of MRF is accurate but time-consuming to calculate; in contrast, approximations to the actual likelihood, say pseudo-likelihood, are efficient to calculate but inaccurate. Thus, how to achieve both accuracy and efficiency simultaneously remains a challenge. In this study, we present such an approach (called clmDCA) for contact prediction. Unlike plmDCA using pseudo-likelihood, i.e., the product of conditional probability of individual residues, our approach uses composite-likelihood, i.e., the product of conditional probability of all residue pairs. Composite likelihood has been theoretically proved as a better approximation to the actual likelihood function than pseudo-likelihood. Meanwhile, composite likelihood is still efficient to maximize, thus ensuring the efficiency of clmDCA. We present comprehensive experiments on popular benchmark datasets, including PSICOV dataset and CASP-11 dataset, to show that: $i)$ clmDCA alone outperforms the existing MRF-based approaches in prediction accuracy.  $ii)$ When equipped with deep learning technique for refinement, the prediction accuracy of clmDCA was further significantly improved, suggesting the suitability of clmDCA for subsequent refinement procedure. We further present successful application of the predicted contacts to accurately build tertiary structures for proteins in the PSICOV dataset. 

\text{Accessibility: } The software clmDCA and a server are publicly accessible through \href{http://protein.ict.ac.cn/clmDCA/}{http://protein.ict.ac.cn/clmDCA/}.
\end{abstract}

\maketitle

\section{Introduction}
In the natural environment, proteins tend to adopt specific tertiary structural conformations (called {\it native structures}) that are solely determined by their amino acid sequences \cite{anfinsen1972studies}. The native structures are stablized by local and global interactions among residues, forming inter-residue contacts with close proximity \cite{gromiha2004inter}. Thus, accurate prediction of inter-residue contacts could provide distance information among residues and thereafter facilitate both free modeling \cite{wu2011improving,marks2012protein,michel2014pconsfold} and template-based modeling approaches \cite{ma2014mrfalign} to protein structure prediction.


A great variety of studies have been conducted for predicting inter-residue contacts, which fall into two categories, namely,  supervised learning approaches  and purely-sequence-based approaches. Supervised learning approaches \cite{di2012deep,eickholt2012predicting,wang2013predicting,Skwark2014ImprovedCP} use training sets composed of residue pairs and contact labels indicating whether these residue pairs form contact or not. Over training sets, machine learning algorithms learn the dependency between contact labels and  features of residue pairs, including sequence profile, secondary structure, solvent accessibility.  The widely-used machine learning algorithms include neural networks, support vector machines, and linear regression models \cite{Fariselli2001PredictionOD, Hamilton2004ProteinCP, MacCallum2004StripedSA, Martin2005UsingIT, Pollastri2002ImprovingTP,Punta2005PROFconNP, Shao2003PredictingIC, Xue2009PredictingRC,Yuan2005BetterPO, Horner2008CorrelatedSA,jones2012psicov, Liu2015ApplicationOL}. Recently, Wang et al. applied deep learning technique to denoise predicted inter-residue contacts, and successfully used predicted contacts to build tertiary structures of several membrane proteins \cite{Wang2017FoldingMP}. 

Unlike the supervised learning approaches, the purely-sequence-based approaches \cite{chiu1991inferring, dunn2008mutual, morcos2011direct,de2013emerging} do not require any training set that contains known contact labels. Instead, the purely-sequence-based approaches begin with collecting homologous proteins of  query protein and constructing multiple-sequence alignment (MSA) of these homologous proteins. Subsequently, coupling columns in MSA are identified to infer contacts among corresponding residues \cite{shindyalov1994can,gobel1994correlated}. The underlying principle lies in the fact that protein structures show considerable conservation during evolutionary process; thus, residues in contact tend to co-evolve to maintain stability of protein structures. Consider two residues being in contact: should one residue mutate and  perturb local structural environment surrounding it, its partner would be more likely to mutate into a physicochemically complementary residue to maintain the whole structure. Thus, co-evolving residue pairs, shown as coupling columns in MSA, are high-quality candidates of residues in contacts. 


The co-evolution analysis strategy, if considering each residue pair individually, is usually hindered by the entanglement of direct and indirect couplings generated purely by transitive correlations. To disentangle direct couplings from indirect ones, an effective way is to consider all residue pairs simultaneously using a unified    model, e.g., Bayesian network \cite{Burger2010DisentanglingDF}, Gaussian distribution \cite{jones2012psicov, andreatta2013prediction,Ma2015ProteinCP}, network deconvolution \cite{Sun2015ImprovingAO}, and Markov random field \cite{Lapedes}. 
Although the Markov random field technique could perfectly model MSA using a joint probability distribution of all residues, maximization of its actual likelihood function is time-consuming as calculating partition function under multiple parameter settings is needed. To overcome this difficulty, a variety of approximation techniques have been proposed as alternatives to likelihood maximization. For example, bpDCA uses message-passing technique to approximate the actual likelihood \cite{Weigt2009IdentificationOD}; mfDCA employs mean field approximation \cite{morcos2011direct} and successfully uses the predicted contacts in {\it de novo} protein structure prediction, and plmDCA completely avoids the calculation of partition function by using pseudo-likelihood as approximation to the actual likelihood and outperforms mfDCA in prediction accuracy  \cite{ekeberg2013improved,kamisetty2013assessing}. 

There is a dilemma in MRF-based approaches to contact prediction: the actual likelihood function of MRF model is accurate but time-consuming to calculate; in contrast, its approximations, say pseudo-likelihood used by plmDCA, are usually efficient to calculate but inaccurate. Thus, how to achieve both accuracy and efficiency simultaneously remains a challenge to the prediction of inter-residue contacts. 

In this study, we present such an approach that achieves both accuracy and efficiency simultaneously.  Unlike plmDCA applying pseudo-likelihood to approximate the actual likelihood function, our approach applied composite likelihood maximization for direct coupling analysis and was therefore named as clmDCA.  Pseudo-likelihood uses the product of conditional probability of individual residues whereas composite likelihood uses the product of conditional probability of all residue pairs and thus is more consistent with the objective of predicting inter-residue contacts. On one side, composite likelihood has been theoretically proved as a better approximation to the actual likelihood function than pseudo-likelihood. On the other side, composite likelihood is still efficient to maximize, which ensures the efficiency of clmDCA. We also investigated the compatibility of clmDCA with subsequent refinement procedure using the deep neural network technique. 

We present comprehensive experiments on popular benchmark datasets, including PSICOV dataset and CASP-11 dataset. Experimental results suggested that: $i)$ clmDCA alone outperforms the existing purely-sequence-based approaches in prediction accuracy.  $ii)$ When enhanced with deep learning technique for denoising, the prediction accuracy of clmDCA was further significantly improved. Compared with plmDCA, clmDCA is more suitable for subsequent refinement by deep learning. We further successfully applied the predicted contacts to accurately build structures of proteins in the PSICOV dataset.

\section{Methods}

\begin{figure}[h!]
	\centerline{\includegraphics[width=\textwidth]{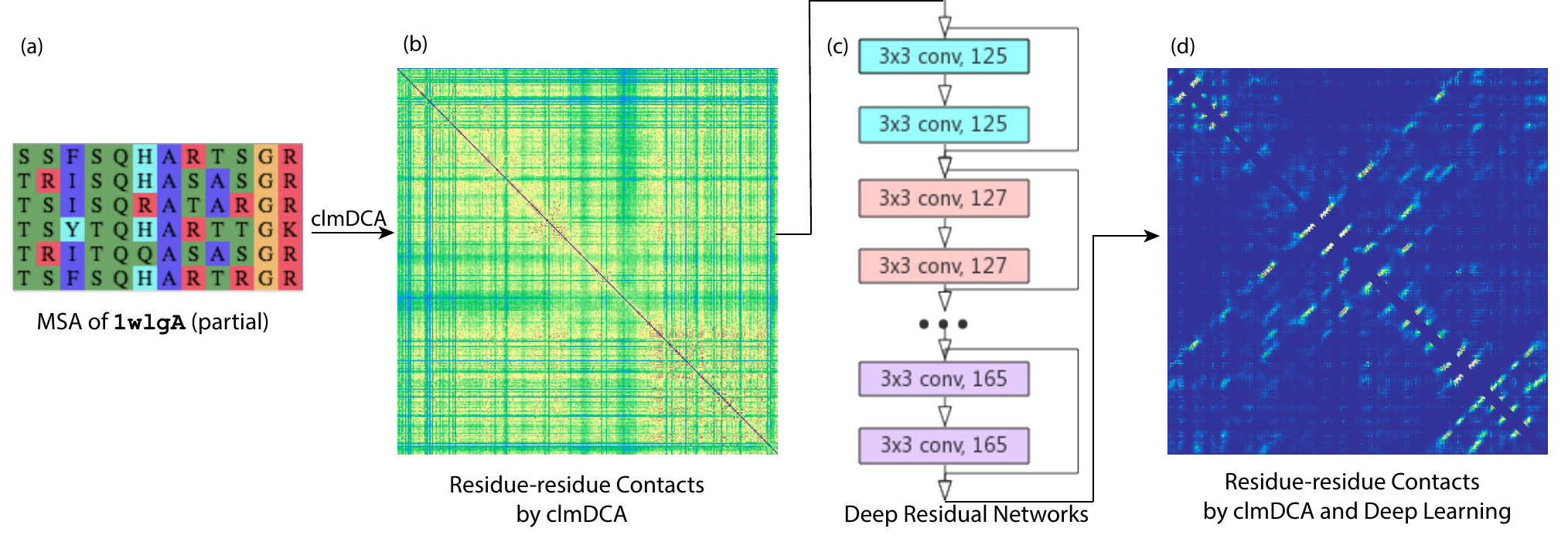}}
	\caption{Procedure of clmDCA to predict inter-residue contacts. (a) For a query protein ({\tt 1wlg\_A} as an example), we identified its homologues by running HHblits \cite{Remmert2011HHblitsLI} against {\it nr90} sequence database (parameter setting: $j: 3, id: 90, cov: 70$) and constructed multiple sequence alignment of these proteins. (b) The correlation among residues in MSA was disentangled using composite likelihood maximization technique, generating prediction of inter-residue contacts. (c) The predicted contacts were fed into a deep neural network for refinement. (d) The  refined prediction of inter-residue contacts. }
	\label{clmDCA}
\end{figure}

For a query protein, clmDCA predicts its inter-residue contacts through the following three steps (Fig. \ref{clmDCA}). First, we construct multiple sequence alignment (MSA) for homologous proteins of the query protein. According to the MSA, the correlations among residues are disentangled using the composite likelihood maximization technique, and are subsequently explored to infer contacts among residues. The generated inter-residue contacts are further refined using a deep residual network. These steps are described in more details as follows. 

\subsection{Modeling MSA using Markov random field} 
For a query protein of length $L$, we denote an MSA of its homologous proteins as  $\{x^{m}\}_{m=1}^{M}$, where $M$ denotes the number of homologous proteins, and  $x^m=(x^m_1, x^m_2, ..., x^m_L)$ represents the $m$-th protein sequence in the MSA. Each element $x^m_i, i=1,2, ..., L$, has a total of 21 possible values, representing 20 ordinary amino acid types and gap in alignment (For the sake of simplicity, we treat gap as a special amino acid type). 

We use a vector of variables $X=(X_{1}, X_{2}, \cdots, X_{L})$ to represent a protein sequence in MSA with $X_i$ representing position $i$ of MSA. According to the maximum entropy principle \cite{Lapedes}, the probability that  $X$ takes a specific value $x^m$ can be represented using Markov random field model \cite{morcos2011direct}: 
\begin{equation}
P(X=x^m) = \frac{1}{Z^{m}} \exp\{\sum_{i=1}^{L}h_{i}(x_{i}^m) + \sum_{i=1}^{L}\sum_{j=i+1}^{L} e_{ij}(x_{i}^m, x_{j}^m)\}
\end{equation}
Here the singleton term $h_{i}(a)$ encodes the propensity for amino acid type $a$ to appear at position $i$, whereas the doubleton term $e_{i,j}(a, b)$ encodes the coupling strength between position $i$ and $j$ when amino acid types $a$ and $b$ appear at these positions, respectively. $Z^{m}$ denotes a partition function acting as a global normalizer to ensure the probabilities of all possible values of $X$ sum to 1. 

The optimal parameters  $h_{i}(a)$ and $e_{i,j}(a, b)$ can be solved via maximizing the likelihood (in logarithm) of all homologous proteins in the MSA, i.e., 
\begin{equation}
\mathcal{L} = \frac{1}{M} \sum_{m=1}^{M}\log P(X=x^{m})
\end{equation}

Finally, we calculated the coupling strength between position $i$  and $j$ using Frobenius  form \cite{jones2012psicov} of the matrix $e_{ij}$:  
\begin{equation}
J_{ij} = \left( \sum_{a=1}^{21}\sum_{b=1}^{21} e_{ij}^2(a, b) \right) ^{\tfrac{1}{2}},
\end{equation} 
which was used to measure the possibility for the corresponding residues of the query protein being in contact.


%
%
%
%

\subsection{Direct coupling analysis using composite likelihood maximization} 
The maximization of the actual likelihood of MRF model is inefficient since the calculation of partition function $Z^{m}$ under multiple parameter settings is needed \cite{Weigt2009IdentificationOD,morcos2011direct}. To circumvent this difficulty,  pseudo-likelihood was used as an approximation to the actual  likelihood $\mathcal{L}$  \cite{ekeberg2013improved,kamisetty2013assessing}:  
\begin{equation}
\mathcal{PL} = \frac{1}{M} \sum_{m=1}^{M} \sum_{i=1}^{L} \log P(X_i=x^m_{i} | X_{\neg i}=x^m_{\neg i})
\end{equation}
Here $P(X_i=x^m_{i} | X_{\neg i}=x^m_{\neg i})$ represents the conditional probability for amino acid type $x^m_{i}$ appearing at position $i$ given the other positions' value $x^m_{\neg i}$. Unlike the actual likelihood $\mathcal{L}$, the approximation $\mathcal{PL}$ is easy to maximize; however,  the deviation between $\mathcal{L}$ and $\mathcal{PL}$ is large, causing inaccurate estimation of parameters in $e_{ij}$ and thereafter inaccurate prediction of inter-residue contacts. 

To better approximate the actual likelihood $\mathcal{L}$, we use composite likelihood $\mathcal{CL}$ instead of  pseudo-likelihood $\mathcal{PL}$  \cite{besag1974spatial}. The composite likelihood is defined as:
\begin{equation}
\mathcal{CL} = \frac{1}{M}\sum_{m=1}^{M} \sum_{c\in C} \log P(X_{c} = x_c^{m} | X_{\neg {c}} = x_{\neg {c}}^{m})
\end{equation}
Here $C$ denotes subsets of variables. This way,  the correlations among all variables within each subset in $C$ are taken into account by $\mathcal{CL}$. 

It should be pointed out that  composite likelihood is a general model with $\mathcal{L}$  and  $\mathcal{PL}$ as its special cases. In particular, when setting $C=\{\{1, 2, \cdots,  L\}\}$, composite likelihood $\mathcal{CL}$ degenerates to the actual likelihood $\mathcal{L}$. On the contrary, when setting $C=\{ \{1\}, \{2\}, \cdots,  \{L\}\}$, the composite likelihood $\mathcal{CL}$ reduces into the pseudo-likelihood $\mathcal{PL}$. 

To match our objective of predicting inter-residue contacts, we set $C$ as all possible residue pairs, i.e.,  ${C}=\{\{1,2\},\{1,3\},\cdots,\{i,j\}, \cdots, \{L-1,L\}\}$. This way, the actual likelihood is approximated using pairwise composite likelihood, which explicitly represents conditional probabilities of all residue pairs as below. 
\begin{align}
\mathcal{CL}_{\text{pairwise}} &=\frac{1}{M}\sum_{m=1}^{M}\sum_{i=1}^{L}\sum_{j>i}^{L}\log P(X_{i,j}=x_{i,j}^{m}|X_{\neg\{i,j\}}=x_{\neg\{i,j\}}^{m}) \nonumber\\
&=\frac{1}{M}\sum_{m=1}^{M}\sum_{i=1}^{L}\sum_{j>i}^{L}\log \frac{1}{Z_{ij}^{m}}\exp\{ h_{i}(x_{i}^{m}) + h_{j}(x_{j}^{m}) +e_{ij}(x_{i}^{m}, x_{j}^{m}) \nonumber\\
&+ \sum_{k\neq i, k \neq j}\left[e_{ik}(x_{i}^{m}, x_{k}^{m}) + e_{jk}(x_{j}^{m}, x_{k}^{m})\right]\}
\end{align}
in which $Z_{ij}$ is a partition function. 
To find optimal parameters $h_{i}$ and $e_{ij}$ such that $\mathcal{CL}_{\text{pairwise}}$ is maximized, we employed the classical Broyden-Fletcher-Goldfarb-Shanno algorithm with efficient calculation of gradients (See Supplementary Material for details). 

The advantages of pairwise composite likelihood technique are two-folds: $i)$ Compared with pseudo-likelihood, pairwise composite likelihood is a better approximation to the actual likelihood. To be more precisely, it has been proved that under any specific parameter setting, $\mathcal{PL} \leq \mathcal{CL}_{\text{pairwise}} \leq  \mathcal{L}$ \cite{yasuda2012}. $ii)$ The gradients of  $\mathcal{CL}_{\text{pairwise}}$  can be calculated in polynomial time. Thus, the pairwise composite likelihood approach achieves both accuracy and efficiency simultaneously. 

\subsection{Refining inter-residue contacts using deep residual network} 
The MRF-based approaches, even being enhanced with direct coupling analysis technique, usually show limited prediction accuracy as they explores MSA of the query protein only but never considers known contacts of other proteins for reference. Recent progresses suggested that this limitation could be effectively avoided by integrating MRF-based approaches with  supervised learning approaches, especially  deep neural networks \cite{jones2014metapsicov, wang2017accurate, liu2017enhancing, Wang2017AnalysisOD}. The power of this integration strategy is rooted in the complementary properties between these two types of approaches: $i)$ The MRF technique considers inter-residue contacts individually but never consider the interdependency among contacts, say {\it clustering pattern} of contacts existing in $\beta$ sheets. $ii)$ In contrast, deep neural networks could learn such contact patterns from known contacts of proteins in training sets, which could be exploited to identify and therefore filter out erroneous predictions by MRF-based approaches.

To refine the predicted contacts by clmDCA, we fed them into a deep residual network \cite{wang2017accurate} for denoising. Deep residual network has its advantages in the ease of  training process and the capacity of considerably deep architecture as each layer learns a residual function with reference to the layer input rather than unreferenced functions\cite{He2015DeepResidualNetwork}. Here, we use a total of 42 convolution layers, organized into 21 residual blocks (Fig. \ref{clmDCA}.c). The convolutional layers have $3\times 3$ filters, each filter aiming to calculate the possibility of a contact between residues $i$ and $j$ according to possibilities of surrounding contacts. The final layer is $softmax$ that transforms the final predicted possibility into the range $[0, 1]$.  
As performed in Ref. \cite{Wang2017FoldingMP}, we also considered the 1D information of the query protein, including sequence profile, predicted secondary structure, solvent accessibility.

\section{Results and discussions}

In our experiments, we tested clmDCA on PSICOV  \cite{jones2012psicov}  data set (containing 150 proteins) and CASP-11 data set (containing 85 proteins). To train the deep residual network for refinement, we constructed a training set through selecting a subset (protein sequence length $< 350$ AA) from the training set used in Ref. \cite{wang2017accurate}. To avoid possible overlap between training set and testing sets, we filtered out the similar proteins shared by training set and test sets. The  criterion of similarity was set as  sequence identity over $25\%$, which has been widely used in previous studies \cite{Zhang2016ImprovingRC,wang2013predicting,Ma2015ProteinCP}. After this filtering operation, the training set contains 3705 proteins in total (available through \href{http://protein.ict.ac.cn/clmDCA/ContactsDeepTraining.tar}{http://protein.ict.ac.cn/clmDCA/ContactsDeepTraining.tar}).  

For each protein in training and test sets, true contacts have been annotated between two residues with a $C_\beta-C_\beta$ ($C_\alpha$ in the case of {\tt Glycine} residues) distance of less than $8 \angstrom$. The performance of contact prediction was evaluated using the mean prediction precision (also known as accuracy), i.e., the fraction of predicted contacts are true \cite{Zhang2016ImprovingRC,morcos2011direct,Ma2015ProteinCP,kamisetty2013assessing,jones2012psicov}. 

In the following subsections, we first evaluated clmDCA and compared it with state-of-the-art MRF-based approaches. Next we examined the difference between clmDCA and plmDCA using protein {\tt 1ne2A} as a concrete example. Then we investigated the enhancement of clmDCA and plmDCA by incorporating the deep learning technique. Finally we presented the application of predicted inter-residue contacts for prediction of protein 3D structures.

\subsection{Overall performance on PSICOV  and CASP-11 datasets}

\begin{table}[h!]
\small
\caption{Contact prediction accuracy on PSICOV benchmark.}
\label{tab:psicov}
\begin{tabular}{ c c c c c c c c c c}
\hline
 Methods  &  \multicolumn{4}{c}{$separation  \geq 6$} & & \multicolumn{4}{c}{$separation  \geq 23$} \\%
 \cline{2-5} \cline{7-10} 
 & $L/10$ & $L/5$ &  $L/2$ & $L$& & $L/10$ & $L/5$ & $L/2$ & $L$\\
\hline
PSICOV  & 0.77 & 0.72 & 0.58 & 0.44 & & 0.72 & 0.64 & 0.47 & 0.34\\
mfDCA    & 0.73 & 0.67 & 0.57 & 0.44 & & 0.71 & 0.64 & 0.49 & 0.36\\
plmDCA  & 0.81 & 0.77 & 0.66 & 0.51 & & 0.78 & 0.71 & 0.56 & 0.40\\
clmDCA   & \text{0.83} & \text{0.80} & \text{0.70} & \text{0.55} && \text{0.81}  & \text{0.75} & \text{0.61} & \text{0.45}\\
plmDCA+DL & 0.92 & 0.90 & 0.85 & 0.75 && 0.89 & 0.86 & 0.74 & 0.59 \\
clmDCA+DL & 0.94 & 0.92 & 0.86 & 0.77 && 0.91 & 0.86 & 0.76 & 0.61 \\
\hline
\end{tabular}
\end{table}

\begin{table}[h!]
\small
\caption{Contact prediction accuracy on CASP-11 targets.}
\label{tab:casp11}
\begin{tabular}{ c c c c c c c c c c}
\hline
 Methods  &  \multicolumn{4}{c}{$separation  \geq 6$} &  & \multicolumn{4}{c}{$separation  \geq 23$} \\%
 \cline{2-5} \cline{7-10} 
   & $L/10$ & $L/5$ &  $L/2$ & $L$& & $L/10$ & $L/5$ & $L/2$ & $L$\\ 
 \hline
PSICOV   & 0.54 & 0.48 & 0.39 & 0.31 && 0.49& 0.43&0.33 &0.24 \\
mfDCA     & 0.49 & 0.44 & 0.37 & 0.30 && 0.48& 0.42&0.33 &0.25  \\
plmDCA   & 0.54 & 0.49 & 0.41 & 0.33  && 0.51& 0.45&0.36&0.26 \\
clmDCA  &\text{0.57} & \text{0.53} &\text{0.44} & \text{0.36}  && \text{0.53}  & \text{0.49} & \text{0.38} & \text{0.29}  \\ 
plmDCA + DL & 0.77 & 0.71 & 0.60 & 0.48 && 0.50 & 0.46 & 0.38 & 0.29  \\ 
clmDCA + DL & 0.86 & 0.81 & 0.72 & 0.60 && 0.69 & 0.64 & 0.52 & 0.40  \\ 
\hline
\end{tabular}
\end{table}

Table \ref{tab:psicov} summarizes the performance of clmDCA, plmDCA, PSICOV and mfDCA on the PSICOV dataset. Following the contact prediction conventions, we filtered out short distance contacts under two settings of sequence separation thresholds ($6$ AA and $23$ AA), and reported the accuracy of top $L/10$, $L/5$, $L/2$, and $L$ predicted contacts. 

\begin{figure}[!h]
\begin{minipage}{1\textwidth}
	\centerline{\includegraphics[width=0.8\textwidth]{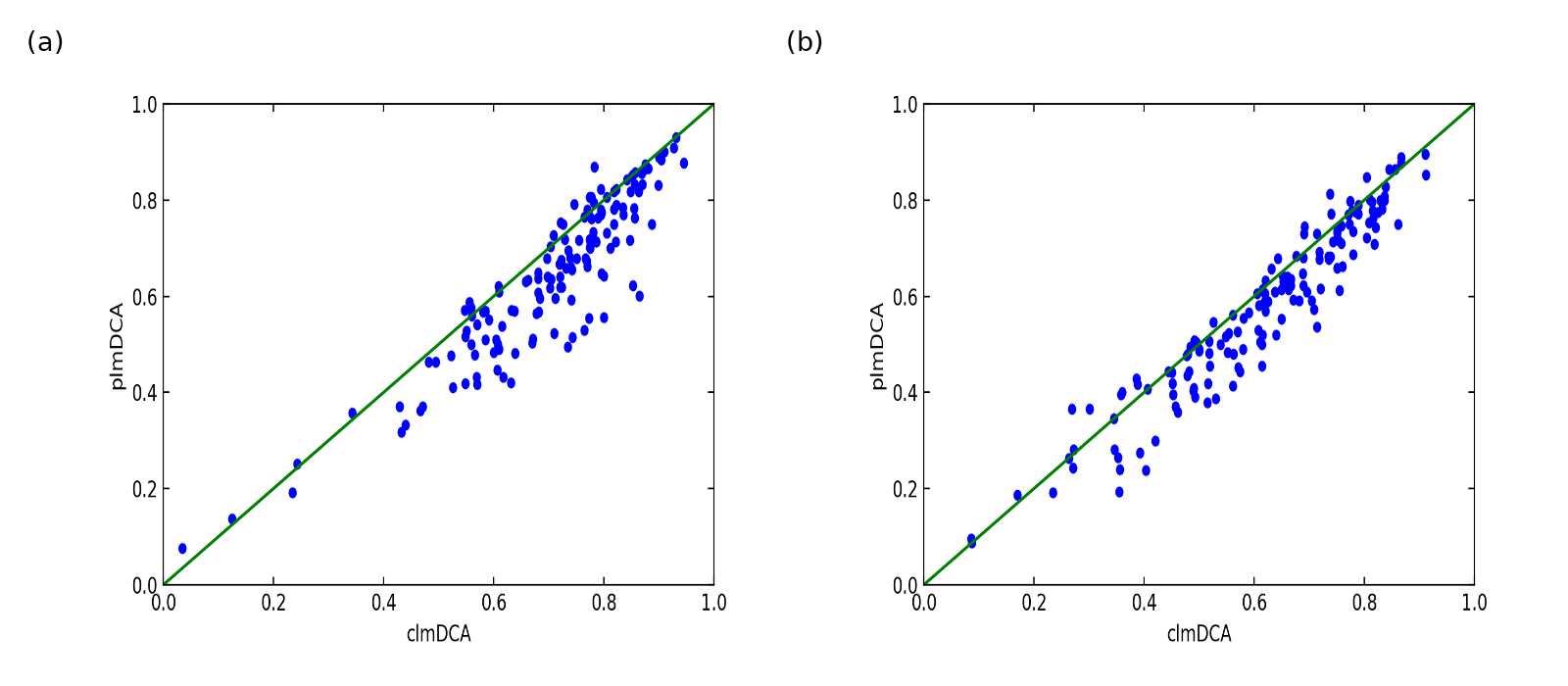}}
\end{minipage}
\caption{Comparison of prediction accuracy of top $L/2$ contacts reported by plmDCA($y$-axis) and clmDCA($x$-axis) with two sequence separation threshold on the PSICOV dataset. (a) Sequence separation $>6$ AA. (b) Sequence separation $>23$ AA.} 
\label{fig:psicov_acc623}
\end{figure}

As shown in Table \ref{tab:psicov} and Figure \ref{fig:psicov_acc623},  clmDCA outperforms plmDCA and other purely-sequence-based approaches.  Take top $L/10$ predictions with the sequence separation threshold $6AA$ as an example. clmDCA achieved prediction precision of $0.83$, which is higher than plmDCA ($0.81$), mfDCA (0.73) and PSICOV ($0.77$). 

Table \ref{tab:casp11} shows that  on the CASP-11 dataset, the prediction accuracy of all these approaches are relatively low than those on the PSICOV dataset. This might be attributed to the difference in MSA quality: the median number of non-redundant homologous proteins is 2374 for proteins in  PSICOV dataset, which is substantially higher than that in CASP-11 dataset (352 homologous proteins on average); the analysis of the effect of the number of effective homologous proteins is shown in the following section. This table suggested that even if the MSA quality is low, clmDCA still outperformed other approaches. 

These tables also suggest that when equipped with deep learning technique for refinement, both plmDCA and clmDCA achieved better prediction accuracy. For example, on the CASP-11 dataset, plmDCA and clmDCA  alone achieved prediction accuracy of only 0.54 and 0.57, respectively (sequence separation $> 6AA$; top $L/10$ contacts). In contrast, by applying the deep learning technique for refinement, the prediction accuracies  significantly increased to 0.77 and 0.86, respectively. More importantly, the improvement of clmDCA (from 0.57 to 0.86) is considerably higher than that of plmDCA (from 0.54 to 0.77), suggesting that clmDCA results are more suitable for refinement using deep learning technique.

\subsection{Comparison of plmDCA and clmDCA: a case study} 

\begin{figure}[!h]
\begin{minipage}{1\textwidth}
	\centerline{\includegraphics[width=0.8\textwidth]{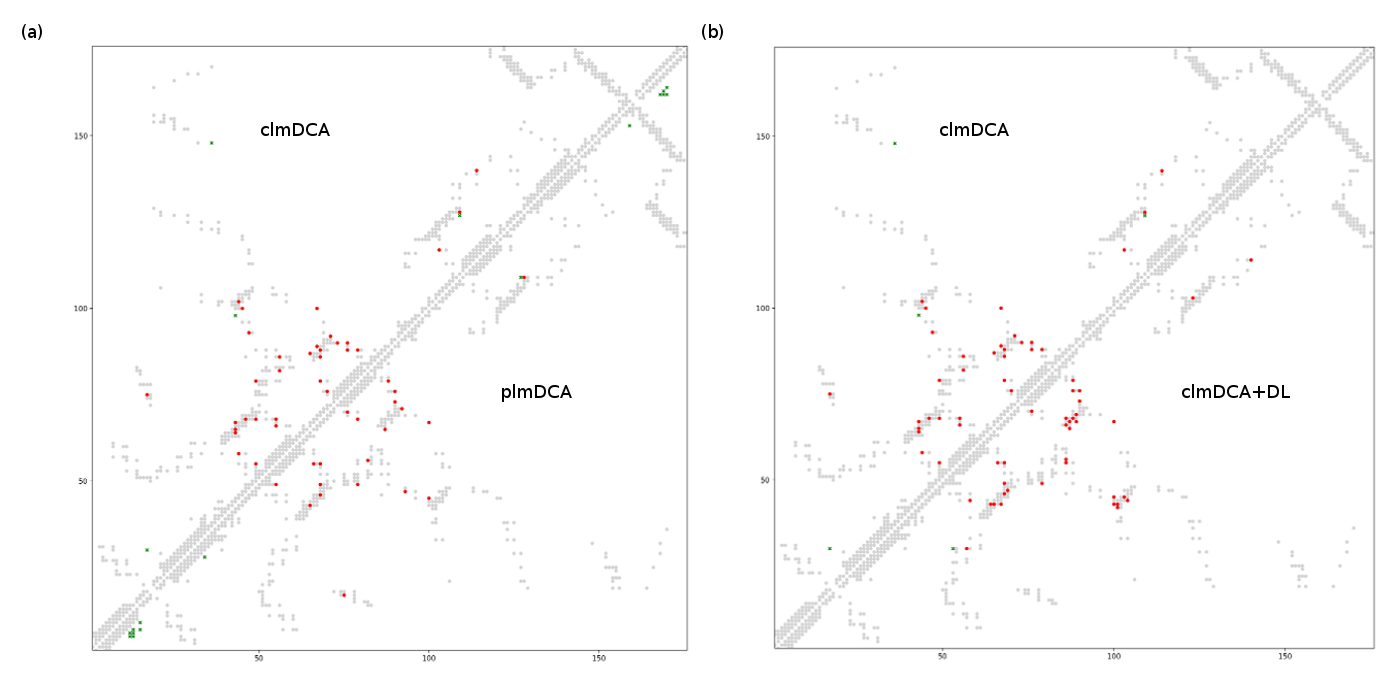}}
\end{minipage}
\caption{Predicted contacts (top $L/5$; sequence separation $>6$ AA) for protein {\tt 1ne2A} by plmDCA and clmDCA. Red (green) dots indicate correct (incorrect) prediction. (a) The comparison between clmDCA (in upper-left triangle) and plmDCA (in
lower-right triangle). (b) The comparison between clmDCA (in upper-left triangle) and clmDCA after refining using deep residual network (in lower-right triangle).}
\label{fig:case-contacts}
\end{figure}

In Figure \ref{fig:case-contacts}, we present the predicted contacts for protein {\tt 1ne2A} by using plmDCA and clmDCA.  By comparing with  true contacts, we observed that clmDCA achieved a contact prediction precision of 0.92, which is significantly higher than plmDCA (prediction precision: 0.50).  

The two approaches, plmDCA and clmDCA, differ only in the way to calculate the parameters $h_{i}$ and $e_{ij}$  and thereafter the coupling strength $J_{ij}$. To reveal this difference, we examined two residue pairs, one  being in contact, and the other  non-contact. As shown in Supplementary Figure  1 (a), the non-contact residue pair {\tt ALA183-ILE189} was incorrectly reported as being in contact by plmDCA (coupling strength: $J_{183, 189} = 1.63$; rank: 14th). In comparison, this pair was ranked 2053th  by clmDCA (coupling strength: $J_{183, 189} = 0.05$) and was not reported as being in contact.

Supplementary Figure 1 (b) shows  {\tt THR75-MSE97} as an example of contacting residue pair. This pair was ranked 40th by plmDCA due to its considerably small  coupling strength   $J_{75, 97} = 1.34$. On the contrary, clmDCA calculated the coupling strength as 0.58  (rank: 12th) and thus correctly reported it as a contact. 
Together these results suggest that compared with plmDCA, clmDCA assigned higher ranks for true contacts.


\subsection{Examining the factors affecting contact prediction} 
The purely-sequence-based approaches use MSA as sole information source; thus, their performance are largely affected by the quality of MSA that is commonly measured using the number of effective homologous proteins (denoted as $N_{eff}$). Most  purely-sequence-based approaches perform perfectly for query protein with high quality, say $N_{eff} \geq 1000$; thus, it is important for a prediction approach to work perfectly when high-quality MSAs are unavailable \cite{Skwark2014ImprovedCP,jones2014metapsicov,jones2012psicov}. 

\begin{figure}[!h]
\begin{minipage}{1\textwidth}
	\centerline{\includegraphics[width=0.8\textwidth]{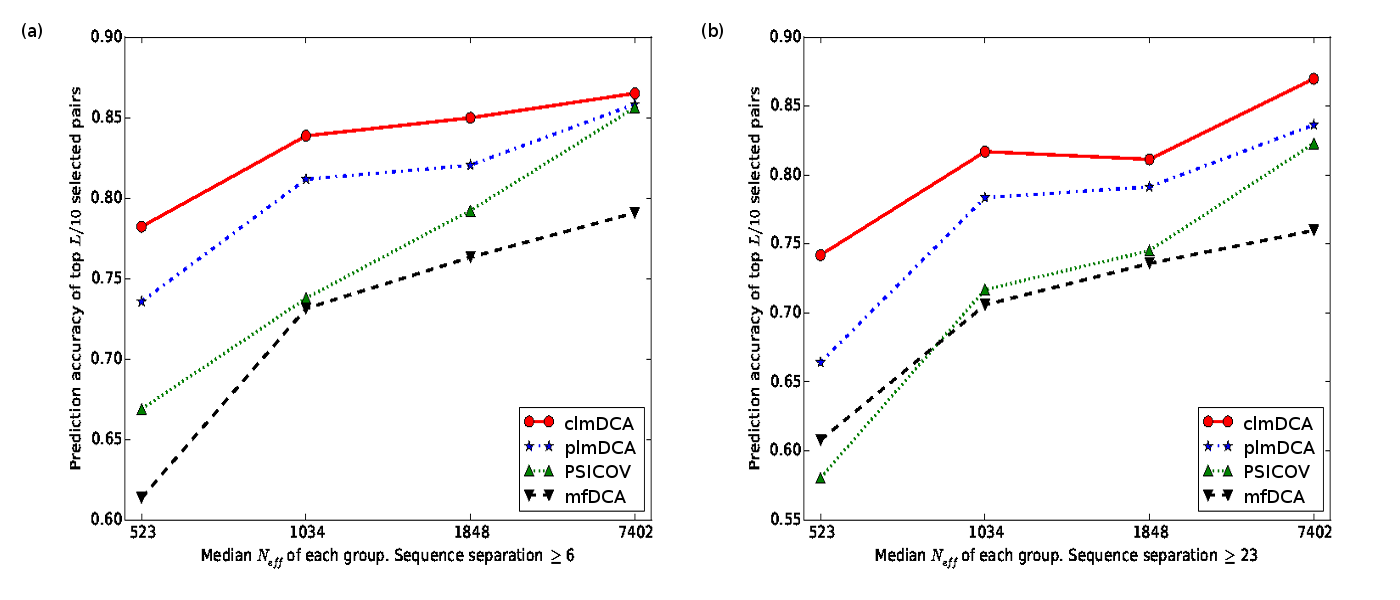}}
\end{minipage}
\caption{The relationship between the prediction accuracy and quality of MSA. Here the quality of MSA is measured using $N_{eff}$, i.e. the number of effective homologous sequences. Dataset: PSICOV. Sequence separation: $>6$ AA} 
\label{fig:psicov_neff_acc}
\end{figure}

Here we examined the affect of $N_{eff}$ on the prediction accuracy of clmDCA. For this end, we divided the proteins in the PSICOV dataset into four groups according to $N_{eff}$ of their MSAs, and calculated the  prediction accuracy for each group individually. As shown in Figure \ref{fig:psicov_neff_acc}, the prediction accuracy of plmDCA, mfDCA, clmDCA and PSICOV increases with $N_{eff}$ as expected. Remarkably, clmDCA outperforms all other approaches even if $N_{eff}$ is only 523, which clearly shows the robustness of clmDCA.

\subsection{Building protein 3D structures using the predicted inter-residue contacts} 

We further applied the predicted inter-residue contacts to build 3D structures of query proteins. For this aim, we run CONFOLD \cite{Adhikari2015CONFOLDRC} with predicted contacts as input. CONFOLD builds protein structure that satisfies the input inter-residue contacts as well as possible. Previous studies have shown that knowing only a few true contacts is sufficient for building high-quality 3D structures \cite{Ovchinnikov2016ImprovedDN}. 

Supplementary Figure 2 compares the quality of structures built using top $L$ contacts predicted by plmDCA, clmDCA alone, and clmDCA together with deep learning. When using contacts predicted by clmDCA alone, the quality of built structures are the same to those built using contacts by plmDCA; however, the combination of clmDCA and deep learning technique showed substantial advantage. Specifically, when using top $L$ contacts predicted by plmDCA as input, we successfully built high-quality structures for 77 proteins in the PSICOV dataset (TMscore $> 0.6$). In contrast, we built high-quality structures for 78 proteins when using predicted contacts by clmDCA. By enhancing clmDCA with deep learning technique, the number of high-quality predictions further increased to 80. 

\begin{figure}[!h]
\begin{minipage}{1\textwidth}
	\centerline{\includegraphics[width=0.6\textwidth]{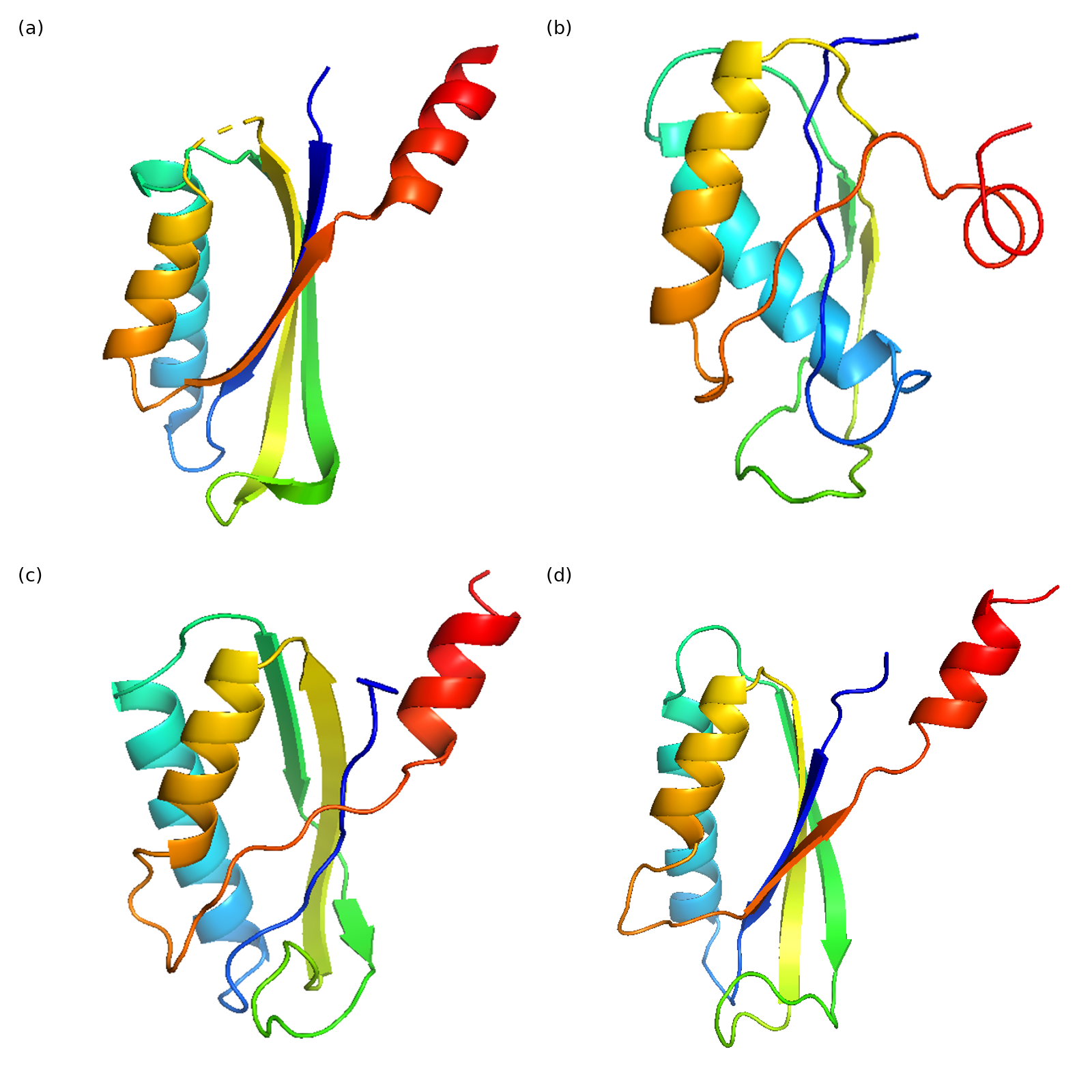}}
\end{minipage}
\caption{
Native structure and predicted structures  for protein {\tt 1vmbA}. (a) Native structure. (b) Structure  built using contacts predicted by plmDCA (TMscore: 0.42).  
(c) Structure  built using contacts predicted by clmDCA alone (TMscore: 0.55). 
(d) Structure  built using contacts predicted by clmDCA together with deep learning for refinement (TMscore: 0.72). 
} 
\label{fig:predicted3d}
\end{figure}

A concrete example is shown in Figure \ref{fig:predicted3d}: For protein {\tt 1vmbA}, the predicted structure has just medium quality (TMscore: 0.55) when using predicted contacts by clmDCA alone. In contrast, when using the refined contacts, the quality of predicted protein structure increased to 0.72. These results clearly demonstrate the effectivity of clmDCA, especially when equipped with deep learning technique, in predicting 3D structures. 
 




\section{Conclusion}

In this study, we present an approach to prediction of inter-residue contacts based on composite-likelihood maximization. Like pseudo-likelihood, composite likelihood is also an approximation to the actual likelihood of Markov random field model and thus avoids the inefficiency in calculating partition function. Compared with pseudo-likelihood, composite likelihood  is much closer to the true likelihood and is more suitable for the subsequent refinement procedure based on deep learning. We present comprehensive results to show that composite-likelihood technique outperforms the existing approaches in terms of prediction accuracy. The predicted contacts were also proved to be useful to predict high-quality structure of query proteins. Together, these results suggest that composite likelihood could achieve both prediction accuracy and efficiency simultaneously. 

We also have tried a hybrid likelihood that combines pseudo-likelihood and the composite likelihood. Experimental results (data not shown here) suggested that this hybrid likelihood achieved prediction accuracy comparable to the application of composite likelihood alone, implying that the correlation information extracted by pseudo-likelihood is nearly completely contained within that extracted by the composite likelihood.

The composite likelihood used in this study is pairwise or 2-order, i.e., we consider the conditional probability of all possible residue pairs. A natural extension is 3- or higher order composite likelihood that considers the conditional  probability of all possible 3 or more residue combinations. Compared with the pairwise composite likelihood, the 3-order composite likelihood showed merely marginal improvement on prediction accuracy but significantly lower efficiency. Thus it is not necessary to apply the 3- or higher order composite likelihood technique in practice.

In this study, we applied the gradient descent technique to maximize composite likelihood. An alternative technique is Gibbs sampling or contrastive divergence, which has been shown in training restricted Boltzmann machine \cite{asuncion2010learning, welling2005learning}. In addition, a generalization of pairwise composite likelihood is tree-reweighted belief propagation \cite{wainwright2003tree}. To further speed up clmDCA, a reasonable strategy is to model residue pairs reported by plmDCA only rather than all possible residue pairs. The implementation of these techniques will be future work of this research. 



\section*{Acknowledgements}

We would like to thank the National Natural Science Foundation of China (31671369 and 31770775) for providing financial supports.

\bibliographystyle{bioinformatics}
{\tiny\bibliography{clmDCA}}


\end{document}